\documentclass[10pt,conference]{IEEEtran}

\IEEEoverridecommandlockouts

\usepackage{cite}
\usepackage{amsmath,amssymb,amsfonts}
\usepackage{algorithmic}
\usepackage{graphicx}
\usepackage{textcomp}
\usepackage{xcolor}
\def\BibTeX{{\rm B\kern-.05em{\sc i\kern-.025em b}\kern-.08em
		T\kern-.1667em\lower.7ex\hbox{E}\kern-.125emX}}

\definecolor{headercolor}{RGB}{0,102,204}   
\definecolor{jsoncolor}{RGB}{153,0,0} 

\begin{document}

    \title{Tool Use as Action: Towards Agentic Control in Mobile Core Networks}
    
	\author{%
		\IEEEauthorblockN{%
			Purna Sai Garigipati\IEEEauthorrefmark{1}\IEEEauthorrefmark{2},
			Onur Ayan\IEEEauthorrefmark{1},
			Kishor Chandra Joshi\IEEEauthorrefmark{2},
			Xueli An\IEEEauthorrefmark{1}}
		\IEEEauthorblockA{\IEEEauthorrefmark{1}Heisenberg Research Center, Huawei Technologies Duesseldorf GmbH,  Munich, Germany\\
			Email: \{purna.sai.garigipati,onur.ayan,xueli.an\}@huawei.com}
		\IEEEauthorblockA{\IEEEauthorrefmark{2}Eindhoven University of Technology, Eindhoven, The Netherlands\\
			Email: \{k.c.joshi\}@tue.nl}
	}
	
	\maketitle
	
\begin{abstract}

    Artificial Intelligence (AI) will play an essential role in 6G. It will fundamentally reshape the network architecture itself and drive major changes in the design of network entities, interfaces, and procedures. The adoption of agentic AI in next-generation networks is expected to enhance network intelligence and autonomy through agents capable of planning, reasoning, and acting, while also opening up new business opportunities. Under this vision, existing network functions are expected to evolve into AI-enabled agents and tools that deliver both connectivity and beyond-connectivity services. As an initial attempt to move toward this vision, this paper presents a tool-based interface design and an experimental prototype that are based on agentic AI for the mobile core network, with the Model Context Protocol (MCP) and the Agent2Agent (A2A) protocol as foundational protocols. MCP is selected to design the interface between the agent and network tools, and the A2A protocol is used for message exchange between AI agents. In such an experimental setup, we analyze packet-level message flows between the agents, tools, and network functions and break down the latency of end-to-end operations, starting from the prompt injection until the completion of the input task. This work demonstrates how an AI agent-based core network combined with network-specific tools can be utilized in next generation mobile systems to execute intent-based tasks.
\end{abstract}


	\section{Introduction}

    A Large Language Model (LLM) is a class of AI systems trained based on large‑scale text corpora to learn statistical patterns of language, enabling them to perform tasks such as comprehension, reasoning, and generation over natural‑language inputs. Although an LLM produces outputs in response to external prompts and does not possess intrinsic goals or autonomous intent, it has emerged as a foundational technology due to its broad applicability in knowledge processing, automation, and decision‑support across multi-domains. On the other hand, an AI agent utilizes an LLM's reasoning capabilities to pursue specific objectives by incorporating memory, planning, and tool use to complete complex requests. 
    
    A multi‑agent system is needed when a single, centralized entity cannot effectively handle the scale, complexity, or dynamism of environments. A collection of specialized AI agents, each optimized for a distinct function, work cooperatively to execute complex, multi‑stage workflows~\cite{10849561,11232814}. 
    Such agentic workflows are already being deployed in real-world applications, ranging from personal assistants that manage complex schedules to autonomous developer agents that write, test, and debug code. Furthermore, autonomous operations agents are now utilized to monitor and dynamically reconfigure IT and cloud infrastructure~\cite{huang2025agentic}.

    This transition is now reaching the telecommunications sector~\cite{etsi_gr_eni_056,etsi_gr_eni_051,3gpp_S2_2511244,2025aiagents,11162291}. 
    AI agents are anticipated to serve as a core architectural construct in 6G mobile networks, operating as autonomous, AI‑driven software entities that fundamentally shape the design of network and corresponding functionalities. As these agents proliferate and collaborate on complex tasks, interoperability and standardized communication mechanism become critical.
    Two notable standards have gained traction recently: the Agent2Agent (A2A) protocol, which is primarily designed for interaction between AI agents~\cite{a2a2024}, and the Model Context Protocol (MCP), which aims to enable how servers expose their tools and data resources to AI agents~\cite{mcp2024}. Together, A2A and MCP form a unified communication framework that enables consistent interaction across LLM‑based multi‑agent systems.

While protocols like MCP are already driving autonomous operations in the broader software ecosystem~\cite{mcp-servers}, their application in telecommunications remains largely unexplored. This paper bridges the gap by investigating how MCP, together with A2A, can be utilized for control plane signaling and message exchange in an AI agent-based mobile core network. The contributions of this paper are as follows:

\begin{itemize}
    \item \textbf{Investigation of Agentic AI in Mobile Core Networks:} We analyze the requirements for designing the mobile core network based on AI agents and tool-based interfaces.
   
    \item \textbf{Protocol Analysis:} With the help of Wireshark, which is an open-source network protocol analyzer, we analyze the content of agent-to-agent (i.e., A2A protocol) and agent-to-tool (i.e., MCP) messages in an agent-driven mobile core network workflow. Moreover, we provide a latency analysis across the entire task life-cycle.
    \item \textbf{Experimental Implementation:} We present a practical implementation of an AI agent-based core network, where agents autonomously delegate tasks to other agents, identify the right tools, and invoke them based on user intent, without relying on hardcoded logic. We demonstrate the feasibility of this agent-driven approach using the OpenAirInterface (OAI) platform.
    
\end{itemize}
    
The remainder of the paper is organized as follows. Section~\ref{sec:Proposed Architecture} presents a tool-based interface design for network functions. Section~\ref{sec:experimental_evaluation} describes the experimental evaluation on the OAI Core and presents the packet-level analysis of a representative network function monitoring and control query and the corresponding latency results. Section~\ref{sec:Conclusion} concludes the paper and outlines directions for future work.

\section{Tool-Based Interface Design for Network Functions}

\label{sec:Proposed Architecture}

Tools enable AI agents to interact with their environment, execute actions, and obtain the information they need. When it comes to the usage of generative AI beyond text generation, tools play a central role. 
Tools are essential for an agent because they provide actions that let an AI agent move from reasoning to real system operation. Without tools, an agent is limited to analysis and text generation; with tools, it can interact with external systems, trigger workflows, and configure network resources, etc. 

For the agent-to-tool interface, we adopt a client-server architecture in alignment with the MCP protocol. Each agent comprises an AI model, typically an LLM, and operates as a tool client. The model gives the agent the ability to perform high-level cognitive tasks, specifically task reasoning, task delegation, tool selection, tool invocation, and response summarization. When an agent determines that a concrete action is required, it interacts with the tool server to perform network-internal operations on the network infrastructure by calling the available tools at the server. For instance, if the agent-to-tool interface is realized using the MCP protocol, the network agent sends MCP requests (i.e., HTTP requests with JSON-RPC payload) to the MCP server, where the tool is located.

When a network AI agent issues a request to invoke a tool, the corresponding tool is activated and executes its defined operation. In case the tool execution involves a further entity (e.g., network functions, middleware, or hardware), the tool execution requires interaction via the interface between the tool (in the tool server) and that specific entity. In such a setting, the tool is responsible for converting the agent’s requests into interface-specific messages. For example, if the tool is connecting agents with another mobile core network function (NF) that produces a service, the tool invocation triggers service-based interface (SBI) calls towards the targeted NF. In order to execute these calls successfully, the system must address the targeted NF services using the standard resource URI structure, as detailed in Section~\ref{sec:experimental_evaluation}. It is important to mention that the tool execution may not always involve a further entity such as an NF. In such cases, the tool execution is performed locally at the tool server and does not involve further communication.

\begin{figure}[!t]
    \centering
    \includegraphics[width=0.9\columnwidth]{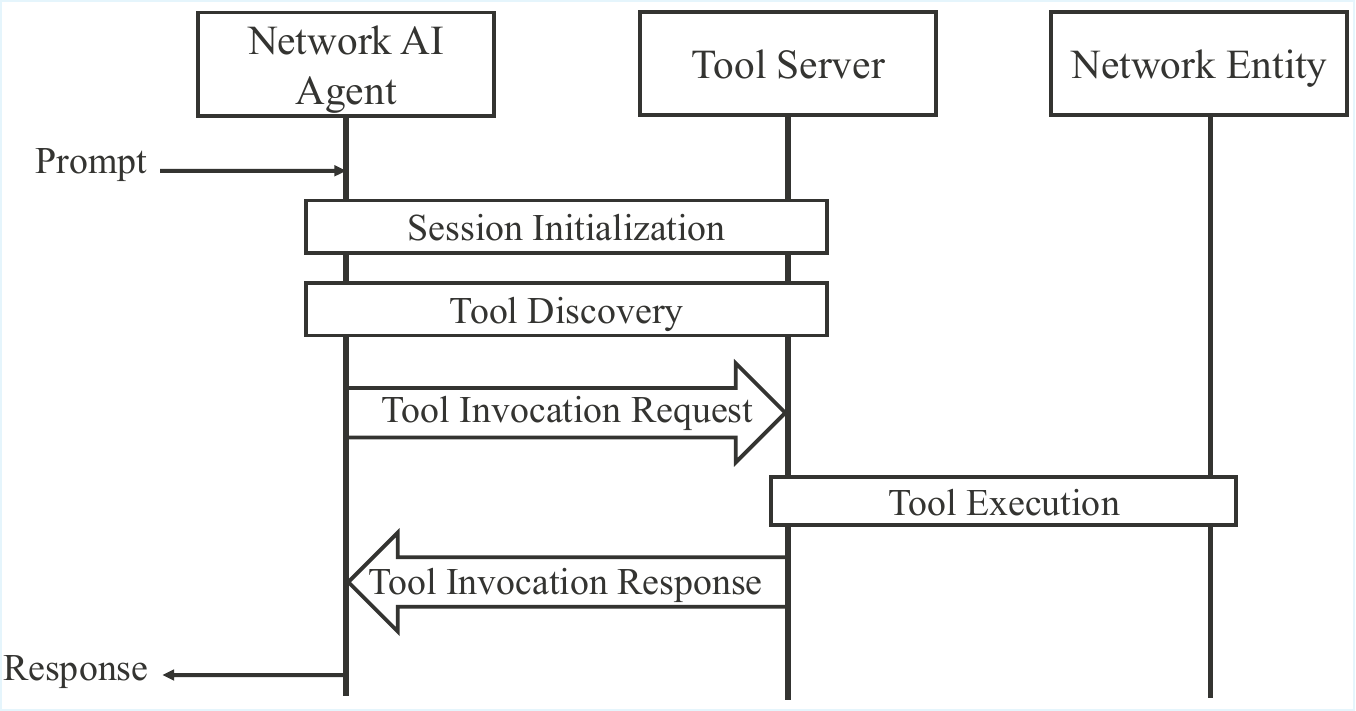}
    \caption{Tool-based interaction flow between a network AI agent and a tool server. In this example the tool execution involves an interaction with a further network entity, which may not always be the case. Example implementations of the network entity are network functions and infrastructure components.}
    \label{fig:mcp_tool_interaction}
     \vspace*{-1mm}
\end{figure}

Fig.~\ref{fig:mcp_tool_interaction} illustrates the logical interaction flow between a network AI agent, a tool server, and a network entity. In this example, the tool server exposes a set of tools. The typical procedure is as follows: The network AI agent first receives a prompt containing a task request, e.g., from a human operator or another AI agent. If an MCP session is not already established, the agent initiates a session with the tool server. Once the session is initialized, the network AI agent requests information about the available tools in order to understand the capabilities exposed by the tool server, including their names, descriptions, input schema, output schema, and metadata. This step corresponds to tool discovery and is crucial before the actual tool invocation. Based on the information provided in the tool discovery response and the original task description, the agent identifies a tool and sends a tool invocation request together with the required input parameters.

\begin{figure*}[!t]
		\centering
		\includegraphics[width=.8\linewidth]{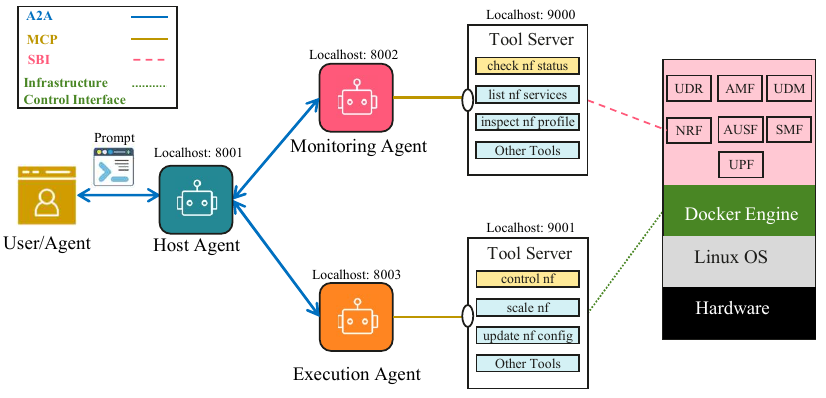}
		\caption{Illustration of our prototype with multiple network AI agents and tool servers. Network AI agents invoke tools to monitor and control further network entities. The highlighted tools in the Tool Server correspond to those invoked during our representative query. The information exchange uses A2A and MCP, while invoked tools trigger SBI calls to NFs or use an infrastructure control interface for Docker-based OAI NF lifecycle operations. In our implementation, the mobile core network includes the Network Repository Function (NRF), Access and Mobility Management Function (AMF), Session Management Function (SMF), Unified Data Management (UDM), Unified Data Repository (UDR), Authentication Server Function (AUSF), and User Plane Function (UPF). }
		\label{fig:agentic_nf_manager}
        \vspace*{-2mm}
\end{figure*}

When the tool is invoked, the tool's software is executed, which corresponds to the tool execution step in Fig.~\ref{fig:mcp_tool_interaction} and involves the interaction with a further network entity in this example. From the agent’s perspective, the tool execution happens transparently, as the AI agent only discovers and invokes these tools, but is agnostic to the tool’s internal implementation. 
If multiple tools are required to complete the task, this is handled by the agent during the planning phase and calls for multiple tool invocations. When the tool execution is completed, the tool invocation response is sent by the tool server to the requesting AI agent, which integrates this result into its reasoning context and forwards the outcome to the user or to another agent for further processing.

In the following section, we focus on the particular architectural example where the tool execution involves interaction with a further entity. As a case study, we take an OAI-based mobile core network implementation and design customized tools that the agent can use to interact with the network functions and infrastructure.

\section{Implementation, Evaluation and Analysis}
\label{sec:experimental_evaluation}

The experimental setup is implemented using the Agent Development Kit (ADK) \cite{adk2025}. We deploy the \texttt{mistral-nemo:latest} model locally via Ollama. All components are deployed locally, enabling full control over execution and precise packet-level observation.

The prototype, illustrated in Fig.~\ref{fig:agentic_nf_manager}, consists of three AI agents interconnected via locally deployed Agent-2-Agent (A2A) endpoints. A \textit{Host Agent} serves as the entry point for user prompts and performs high-level intent interpretation and task delegation. Based on the intent, it delegates tasks to one of the two specialized sub-agents. The \textit{Monitoring Agent} is responsible for monitoring tasks related to NFs, such as checking NF registration status, listing available NF services, and inspecting NF profiles, while the \textit{Execution Agent} handles control and management actions, including starting, stopping, or restarting NFs, as well as updating NF configurations or scaling NFs on demand. In our testbed, each sub-agent embeds an MCP client and is bound to a dedicated MCP server that exposes a set of domain-specific tools.

The mobile core network is implemented via the OAI framework and includes the standard control- and user-plane NFs as illustrated in Fig.~\ref{fig:agentic_nf_manager}. These NFs expose standardized Service-Based Interfaces (SBIs) described using the OpenAPI specification and implemented over HTTP/2 with JSON serialization, as specified in \cite{3gpp23501,ts29501}. NFs register the produced services to the NRF, discover other NFs through the NRF, and invoke their services via SBIs. Each NF service is addressed by a resource URI~\cite{ts29501} of the form \texttt{\{apiRoot\}/<apiName>/}\\%
\texttt{<apiVersion>/<apiSpecificResourceUriPart>}. Here, \texttt{apiRoot} contains the scheme (“http” or “https”) and the authority (host and optional port). The \texttt{<apiName>} identifies the API, for example “nnrf-disc”, and \texttt{<apiVersion>} indicates the version of that API, such as “v1”. These three components together define the base API URI, while \texttt{<apiSpecificResourceUriPart>} identifies a specific resource within that API. 

\subsection{Packet-Level Analysis}
\label{subsec:packet_analysis}

To verify and evaluate the tool-based network control via agents in our prototype, we consider the representative query
\emph{``Check the operational status of the AMF and start it if it is inactive.''} The AMF container is intentionally stopped prior to issuing the query in order to create a controlled test condition.

The user prompt is first received by the Host Agent via the A2A interface. The Host Agent interprets the request as a compound task consisting of (i) network function status inspection and (ii) conditional lifecycle control. Based on this interpretation, the Host Agent delegates the inspection task to the Monitoring Agent. Upon receiving the inspection result, the Host Agent conditionally delegates further tasks to the Execution Agent. Throughout this process, all interactions between the AI agents and tool servers are performed via the MCP protocol. The agents themselves do not directly interact with network entities; all such interactions are mediated by the Tool Server.

Table~\ref{tab:packet_trace} provides a compact, end-to-end view of the packet-level interactions across the A2A, MCP, and SBI interfaces.

\begin{table*}[!t]
\caption{Packet-level trace AMF inspection and control query.}
\label{tab:packet_trace}
\centering
\setlength{\tabcolsep}{15pt}
\renewcommand{\arraystretch}{1.15}
\begin{tabular}{l l l l}
\hline
\textbf{ID} & \textbf{Source $\rightarrow$ Destination} & \textbf{Interface} & \textbf{Purpose} \\
\hline\hline
A1  & User $\rightarrow$ Host Agent 
    & A2A 
    & Submit user prompt to Host Agent \\

A2  & Host Agent $\rightarrow$ Monitoring Agent 
    & A2A 
    & Delegate NF status inspection \\

M1  & Monitoring Agent $\rightarrow$ MCP Server (Localhost: 9000) 
    & MCP 
    & Discover available inspection tools \\

M2  & Monitoring Agent $\rightarrow$ MCP Server (Localhost: 9000) 
    & MCP 
    & Invoke NF status inspection tool \\

S1  & MCP Tool $\rightarrow$ NRF 
    & SBI 
    & Query NF registration state \\

M2' & MCP Server (Localhost: 9000) $\rightarrow$ Monitoring Agent 
    & MCP 
    & Return inspection result (AMF inactive) \\

A3  & Host Agent $\rightarrow$ Execution Agent 
    & A2A 
    & Delegate lifecycle control action \\

M3  & Execution Agent $\rightarrow$ MCP Server (Localhost: 9001) 
    & MCP 
    & Discover lifecycle control tools \\

M4  & Execution Agent $\rightarrow$ MCP Server (Localhost: 9001) 
    & MCP 
    & Invoke NF lifecycle control tool \\

M4' & MCP Server (Localhost: 9001) $\rightarrow$ Execution Agent 
    & MCP 
    & Return lifecycle execution result (AMF started) \\
\hline
\end{tabular}
\vspace*{-4mm}
\end{table*}

\subsubsection*{\textbf{A2A-based Delegation}}

The A2A messages (A1--A3) are used exclusively for task delegation and result propagation between agents and the user, as summarized in Table~\ref{tab:packet_trace}. In step (A1), the user submits the original request to the Host Agent via the A2A interface, where the human operator acts as the client and the Host Agent acts as the server. Upon receiving this prompt, the Host Agent retrieves the agent cards of available peer agents in order to identify their capabilities and corresponding HTTP endpoints.

Based on the inferred intent of the request, the Host Agent selects the Monitoring Agent and issues an A2A delegation request (A2) to initiate the network function inspection task. The destination endpoint of this request is obtained from the Monitoring Agent’s agent card prior to delegation. The excerpt below shows a representative \texttt{message/send} request generated by the Host Agent. In all subsequent excerpts, only essential fields required for task execution are shown; auxiliary headers, message history, and framework-specific metadata are omitted, and color coding is used for clarity, with HTTP headers shown in blue and the HTTP body shown in red.

\begin{quote}\footnotesize
\noindent\textbf{A2A Delegation Request:}\\
{\color{headercolor}\texttt{POST / HTTP/1.1}\\
\texttt{Host: localhost:8002}\\
\texttt{Content-Type: application/json}}\\
{\color{jsoncolor}\texttt{\{"id":"121aace5-78a6-4fce-adfd-93598a60b67b",\\
"jsonrpc":"2.0","method":"message/send",\\
"params":\{"message":\{"kind":"message",\\"role":"user",\\
"parts":[\{"kind":"text","text":\emph{"Check the operational status of the AMF and start it if it is inactive."}\}]\}\}\}}}
\end{quote}

The actual task intent is conveyed through the \texttt{parts} field of the JSON-RPC message body, with the message content encoded in the \texttt{text} subfield, as specified by the A2A protocol. The \texttt{role} field indicates the message direction, where \texttt{"role":"user"} denotes client-to-server communication and \texttt{"role":"agent"} denotes server-to-client communication\footnote{The \texttt{"user"} role is slightly misleading, and it is used to indicate the initiator of an A2A interaction. It does not have to be a human user and it can be an AI agent operating as the client, where the receiving agent is the server.}.

The corresponding response indicates successful completion of the delegated task and returns the outcome to the Host Agent. The result is represented as a completed task containing an agent-originated message that summarizes the execution outcome.

\begin{quote}\footnotesize
\noindent\textbf{A2A Delegation Response:}\\
{\color{headercolor}\texttt{HTTP/1.1 200 OK}\\
\texttt{Content-Type: application/json}}\\
{\color{jsoncolor}\texttt{\{"id":"121aace5-78a6-4fce-adfd-93598a60b67b",\\"jsonrpc":"2.0",\\"result":\{\\
\ \ "kind":"task",\\
\ \ "status":\{"state":"completed",\\
\ \ \ \ "message":\{"role":"agent",\\
\ \ \ \ \ \ "parts":[\{"kind":"text","text":\emph{"It seems like the AMF is currently inactive or not properly registered with the NRF. To address this issue, we'll need to start the AMF."}\}]\}\}\}\}}}
\end{quote}

Following receipt of the inspection result, the Host Agent proceeds in the same manner to delegate the lifecycle control action to the Execution Agent (A3), using the endpoint specified in the Execution Agent’s agent card. This subsequent delegation follows the same A2A interaction pattern shown above, ensuring consistent task propagation across agents.

\begin{table*}[!t]
\caption{LATENCY ANALYSIS OF TOOL-BASED NF MONITORING AND CONTROL}
\label{tab:latency_summary_expanded}
\centering
\setlength{\tabcolsep}{15pt} 
\renewcommand{\arraystretch}{1.15}
\begin{tabular}{l c c c c}
\hline
\textbf{Component} & \textbf{Mean [s]} & \textbf{Standard Deviation [s]} & \textbf{Min [s]} & \textbf{Max [s]} \\
\hline\hline
Host Agent LLM reasoning and delegation planning & 2.35 & 0.01 & 2.35 & 2.36 \\
Monitoring Agent (total) & 4.50 & 0.97 & 3.51 & 6.78 \\
\quad MCP tool listing & 0.02 & $<$0.01 & 0.01 & 0.05 \\
\quad MCP tool call + SBI execution & 0.58 & 0.15 & 0.41 & 0.81 \\
\quad LLM tool selection + result synthesis & 3.11 & 0.90 & 2.31 & 5.42 \\
Execution Agent (total) & 4.99 & 0.49 & 4.26 & 5.60 \\
\quad MCP tool listing & 0.02 & $<$0.01 & 0.02 & 0.03 \\
\quad MCP tool call + system execution & 1.17 & 0.44 & 0.73 & 1.97 \\
\quad LLM tool selection + result synthesis & 3.03 & 0.49 & 2.42 & 3.93 \\
Aggregate A2A Agent Card Retrieval and delegation & $\approx$0.98 & 0.03 & 0.93 & 1.02 \\
\hline
\textbf{End-to-end latency} & \textbf{12.81} & \textbf{1.39} & \textbf{11.08} & \textbf{15.78} \\
\hline
\end{tabular}
\vspace*{-3mm}
\end{table*}

\subsubsection*{\textbf{MCP-Based Inspection via Monitoring Agent}}
After receiving the delegated inspection task from the Host Agent, the Monitoring Agent establishes, if not already present, an MCP session with its associated MCP server and initiates tool discovery by issuing a \texttt{tools/list} request (M1). This step allows the agent to identify the capabilities exposed by the MCP server and to select an appropriate status inspection tool based on the declared schema of the tools. For clarity and brevity, only the relevant inspection tool is shown in the excerpt below. In practice, the response may contain multiple tools, each described by its schema, as illustrated in the excerpt. The excerpt highlights the \texttt{check\_nf\_status} tool, including its required input parameter (\texttt{nf\_type}) and expected output field, which together provide the information necessary for the agent to construct a valid tool invocation in the subsequent step.

\begin{quote}\footnotesize
\noindent\textbf{MCP Tool Listing Request:}\\
{\color{headercolor}\texttt{POST /mcp HTTP/1.1}\\
\texttt{Host: 127.0.0.1:9000}\\
\texttt{Content-Type: application/json}}\\
{\color{jsoncolor}\texttt{\{"jsonrpc":"2.0","method":"tools/list","id":18\}}}
\end{quote}

\begin{quote}\footnotesize
\noindent\textbf{MCP Tool Listing Response:}\\
{\color{headercolor}\texttt{HTTP/1.1 200 OK}\\
\texttt{Content-Type: text/event-stream}}\\
{\color{jsoncolor}\texttt{\{"jsonrpc":"2.0","id":18,\\
"result":\{"tools":[\\
\ \ \{"name":"check\_nf\_status",\\
\ \ \ \ "description":"Checks whether a core network function is currently active (registered in the NRF)",\\
\ \ \ \ "inputSchema":\{\\
\ \ \ \ \ "properties":\{"nf\_type":\{"type":"string"\}\},\\
\ \ \ \ \ "required":["nf\_type"]\},\\
\ \ \ \ "outputSchema":\{\\
\ \ \ \ \ "properties":\{"result":\{"type":"string"\}\},\\
\ \ \ \ \ "required":["result"]\}\}\\
]\}\}}}
\end{quote}

Based on the discovered tool description, the Monitoring Agent selects the \texttt{check\_nf\_status} tool and issues the corresponding MCP tool invocation (M2). The tool call explicitly supplies the network function type as an argument, with \texttt{nf\_type} set to \texttt{AMF}. This parameter value is derived directly from the original user request, which specifies the AMF as the target network function for inspection. The corresponding MCP tool call request and response are shown below.

\begin{quote}\footnotesize
\noindent\textbf{MCP Tool Call Request:}\\
{\color{headercolor}\texttt{POST /mcp HTTP/1.1}\\
\texttt{Host: 127.0.0.1:9000}\\
\texttt{Content-Type: application/json}}\\
{\color{jsoncolor}\texttt{\{"jsonrpc":"2.0","method":"tools/call","id":20,}\\
\texttt{\ \ "params":\{"name":"check\_nf\_status",}\\
\texttt{\ \ \ \ "arguments":\{"nf\_type":"AMF"\}\}\}}}
\end{quote}

\begin{quote}\footnotesize
\noindent\textbf{MCP Tool Call Response:}\\
{\color{headercolor}\texttt{HTTP/1.1 200 OK}\\
\texttt{Content-Type: text/event-stream}}\\
{\color{jsoncolor}\texttt{\{"jsonrpc":"2.0","id":20,\\
"result":\{"structuredContent":\{\\
\ \ "result":\emph{"AMF is not active or not registered in the NRF"}\},\\
"isError":false\}\}}}
\end{quote}

\subsubsection*{\textbf{Underlying SBI Interaction}}

Upon receiving the tool invocation, the MCP server executes the tool, which sends out an SBI request toward the Network Repository Function (NRF) to query AMF status information. This interaction corresponds to step (S1) in Table~\ref{tab:packet_trace}. Specifically, the tool issues an HTTP/2 \texttt{GET} request to the \texttt{/nnrf-nfm/v1/nf-instances} endpoint with the NF type specified as a query parameter, directed to the IP address and port of the target NRF instance,  following the SBI URI structure described earlier.

\begin{quote}\footnotesize
\noindent\textbf{SBI Request (HTTP/2):}\\
{\color{headercolor}\texttt{:method GET}\\
\texttt{:path /nnrf-nfm/v1/nf-instances?nf-type=AMF}\\
\texttt{accept: application/json}}
\end{quote}

\begin{quote}\footnotesize
\noindent\textbf{SBI Response (HTTP/2):}\\
{\color{jsoncolor}\texttt{:status 200}\\
\texttt{content-type: application/json}\\}
{\color{jsoncolor}\texttt{\{"\_links":\{"item":[],"self":""\}\}}}
\end{quote}

The empty \texttt{item} list in the SBI response indicates that no active AMF instance is currently registered in the NRF. If the AMF were active, the response would include the IP address of the corresponding AMF instance. The result is mapped by the MCP server into a structured tool response and returned to the Monitoring Agent (M2’), which forwards it to the Host Agent via A2A.

\subsubsection*{\textbf{Lifecycle Control via Execution Agent}}

Upon confirmation that the AMF is inactive, the Host Agent delegates a corrective lifecycle task to the Execution Agent (A3). The Execution Agent follows the same MCP-based interaction pattern described above for the Monitoring Agent: it retrieves the list of available tools (M3) and selects the appropriate lifecycle control tool based on its declared schema.

The \texttt{control\_nf} tool accepts the target network function identifier and the desired lifecycle action (start, stop, or restart) as input parameters. In this workflow, since the inspection step indicates that the AMF is inactive, the target network function is set to \texttt{AMF} and the lifecycle action is set to \texttt{start}. The operation is requested via the MCP tool invocation (M4). Unlike the inspection tool, this action does not trigger an SBI request; instead, the \texttt{control\_nf} tool uses the infrastructure control interface shown in Fig.~\ref{fig:agentic_nf_manager} to execute the requested lifecycle operation by issuing a Docker command on the host running the OAI core. The execution outcome is returned to the Execution Agent as the corresponding MCP tool response (M4’), and subsequently propagated back to the Host Agent via A2A.

This packet-level analysis grounds the proposed agentic workflow in concrete A2A, MCP, and SBI message exchanges, using representative excerpts from Wireshark traces to illustrate each step of the process. Building on this detailed view, the next subsection examines the latency characteristics of the same workflow and quantifies the time spent across different processing stages.

\subsection{Latency Analysis}
\label{subsec:latency_analysis}

This subsection analyzes the latency characteristics of the agentic NF monitoring and control workflow introduced in the packet-level analysis, using latency measurements obtained from 10 experimental runs. The prompt remains the same for each run. All timings are derived from Wireshark packet timestamps and summarized in Table~\ref{tab:latency_summary_expanded}. The total end-to-end latency of the evaluated workflow has a mean value of 12.81~s, measured from the arrival of the user request at the Host Agent to the delivery of the final response back to the user. Across the runs, the observed end-to-end latency ranges from 11.08~s to 15.78~s, with a standard deviation of 1.39~s. This interval spans multiple agent decision points, task delegations, and tool-driven interactions, providing a complete view of how time is distributed across the agentic control loop.

A first significant latency component arises at the Host Agent during intent interpretation and delegation planning. This phase exhibits a mean latency of 2.35~s with very low variance across runs, indicating stable LLM inference behavior for this reasoning step. During this interval, the Host Agent reasons over the compound request, determines that inspection is required prior to control, and schedules subsequent delegation steps. 

The Monitoring Agent accounts for a mean total execution time of 4.50~s, measured from receipt of the delegated task to propagation of the inspection outcome back to the Host Agent. Within this interval, MCP tool listing is invoked several times by the agent during tool selection, tool invocation, and result synthesis, yet it completes with a mean latency of 0.02~s, confirming that capability discovery introduces negligible overhead. The combined mean latency for the MCP tool call and SBI execution is 0.58~s, while the core SBI HTTP/2 signaling toward the NRF alone accounts for only approximately 0.49~ms. The dominant contributors are instead LLM-bound stages: tool selection based on the advertised schema and result synthesis for reporting the inspection outcome. These LLM-driven stages account for the majority of the monitoring agent’s execution time, with a mean latency of 3.11~s.

A similar pattern is observed for the Execution Agent, whose mean end-to-end execution time is 4.99~s. MCP tool listing again accounts for minimal delay, while the combined MCP tool execution and system-level operation (starting the AMF container) exhibits a mean latency of 1.17~s. The remaining 3.03~s, on average, is spent in LLM-based tool selection and result synthesis. Beyond per-agent execution, timestamp analysis shows that the aggregate A2A-based delegation and response coordination contributes approximately 0.98~s on average to the end-to-end workflow, with limited variation across runs. This time captures all agent-to-agent handovers and result propagation steps combined, including task delegation and final response delivery, and represents only a minor fraction of the total latency.

Overall, the latency analysis reveals a consistent trend across all runs of the workflow. Protocol- and interface-level mechanisms, including A2A messaging and MCP interactions, and SBI-based service access, incur comparatively low and stable overhead, while LLM-driven reasoning stages dominate the response time and contribute most of the observed variability.

\section{Conclusion}
\label{sec:Conclusion}

This paper demonstrates an initial implementation of agentic control for mobile core networks through tool-based interfaces using existing technologies. By combining MCP for tool-based access to network functions with A2A for agent coordination, we implement and evaluate an agent-driven control workflow on an OAI-based core network. Packet-level analysis using Wireshark confirms how user intent is propagated through agents, MCP tools, and core network interfaces, while the latency analysis shows that protocol overhead is small compared to the time required for LLM inference during intent reasoning, delegation planning, tool selection, and result summarization. While our prototype focuses on NF monitoring and lifecycle control tasks, this workflow can be seamlessly extended to a wide variety of use cases by integrating the right set of domain-specific tools.

In future work, we will conduct a deeper investigation into how the mobile core network can be re-designed based on AI agents and tools, an approach that goes beyond conventional service-based architecture. Moreover, we will investigate various mechanisms and design considerations on how to accelerate LLM inference time in order to guarantee telco-grade requirements.

\vspace*{-1mm}
\section*{Acknowledgements}
This work has been supported by the European Union’s Horizon Europe MSCA-DN programme through the SCION Project under Grant Agreement No. 101072375.

\end{document}